# List Factoring and Relative Worst Order Analysis[*][**]

Martin R. Ehmsen[1], Jens S. Kohrt[1,2], and Kim S. Larsen[1]

[1] Department of Mathematics and Computer Science
University of Southern Denmark, Odense, Denmark
{ehmsen,svalle,kslarsen}@imada.sdu.dk
[2] CP$^3$-Origins, University of Southern Denmark, Odense, Denmark

**Abstract** Relative worst order analysis is a supplement or alternative to competitive analysis which has been shown to give results more in accordance with observed behavior of online algorithms for a range of different online problems. The contribution of this paper is twofold. First, it adds the static list accessing problem to the collection of online problems where relative worst order analysis gives better results. Second, and maybe more interesting, it adds the non-trivial supplementary proof technique of list factoring to the theoretical toolbox for relative worst order analysis.

## 1 Introduction

The static list accessing problem [28,4] is a well-known problem in online algorithms. Many deterministic as well as randomized algorithms are known, and these have been investigated theoretically as well as experimentally. See [10] for a discussion of the importance of the problem in relation to dictionary implementation, connections to paging, and applications in compression algorithms. For readers unfamiliar with the standard algorithms for list accessing or relative worst order analysis, we refer to the rigorous definitions in Sections 2 and 3.

The starting point for our work was the discrepancy between the findings obtained using competitive analysis [20,28,22] and the observations made through experimental work. Competitive analysis finds that Move-To-Front is optimal while Frequency-Count and Transpose have lower bounds on their competitive ratio which grow linearly with the length of the list [10]. In contrast, experimental results [9] suggest that Move-To-Front and Frequency-Count are almost equally good and both are far better than Transpose. Results from [7] seem to indicate the same.

To a large extent driven by the paging problem [10] and the difficulties there in theoretically separating various algorithm proposals, many alternative performance measures have been developed to supplement standard competitive analysis. Examples include [30,8,23,24,11,5]; see [16] for a survey. Some of these measures are tailored towards a specific online problem, whereas others are more generally applicable; see [13] for a comparative study of these measures on a simple problem.

Of these alternatives to competitive analysis relative worst order analysis [11,12] is the measure that has been applied to the largest variety of online problems. Results that are in accordance with experiments have been derived for a range of fairly different online problems in situations where competitive analysis has given the "wrong" answer. Online problems of this nature include (but are not limited to) the following:

---

[*] This work was supported in part by the Danish Natural Science Research Council.
[**] Preprint: CP3-Origins-2010-41, IMADA-PP-2010

- For classical bin packing, Worst-Fit is better than Next-Fit [11].
- For dual bin packing, First-Fit is better than Worst-Fit [11].
- For paging, LRU is better than FWF (Flush-When-Full) and look-ahead helps [12].
- For scheduling, minimizing makespan on two related machines, a post-greedy algorithm is better than scheduling all jobs on the fast machine [18].
- For bin coloring [25], a natural greedy-type algorithm is better than just using one open bin at a time [17].
- For proportional price seat reservation, First-Fit is better than Worst-Fit [14].

We apply relative worst order analysis to the static list accessing problem. We first extend the list factoring technique [9,3] known from competitive analysis to relative worst order analysis. We then apply the technique to the three deterministic online list accessing algorithms Move-To-Front, Time-Stamp, and Frequency-Count. We show that these algorithms are equally good and much better than Transpose when analyzed using relative worst order analysis, thereby obtaining results that are in accordance with the cited experimental work.

Adding static list accessing to the collection of problems above where relative worst order analysis gives better or more nuanced results than competitive analysis is a step in documenting to what extent relative worst order analysis is generally applicable. However, we find it more interesting that relative worst order analysis can be equipped with a powerful supplementary proof techniques such as list factoring. To our knowledge, relative worst order analysis is the first of the alternative performance measures to be equipped with a list factoring lemma.

Some of the deterministic list accessing algorithms are quite old. It is difficult to pin-point the origin of Frequency-Count, since it is intimately related to probability theoretical considerations, and it is not clear when it started being viewed as an algorithm. Move-To-Front and Transpose were formulated in [26]. Time-Stamp [1] is a deterministic algorithm that arose as a special case of a family of randomized algorithms.

We also consider the randomized algorithms BIT [27] and Randomized-Move-To-Front[1]. Deterministic and randomized online algorithms are often compared informally, but it is not clear how much sense it makes to compare a worst-case guarantee with an average-case performance. We compare the two randomized algorithms to each other and find them incomparable whereas competitive analysis slightly favors the former (often referred to as a "surprising result"), showing that BIT is $\frac{7}{4}$-competitive and Randomized-Move-To-Front is 2-competitive against an oblivious adversary [27,19].

For early related work, we refer the reader to [4,10]. Newer work obtains separations between list accessing algorithms by analyzing these with respect to some measure of locality of reference [2,6,15].

## 2   List Accessing

In the *static list accessing problem* [28,4], we have a fixed collection of items arranged in a linear list, $\mathcal{L} = (a_1, a_2, \ldots, a_\ell)$, of length $\ell$. The request sequence, I, consists of requests of *access* to items in the list, and the accesses must be served in an online manner. The cost of accessing an item depends on its position (index) in the list. In the *full cost model*, accessing an item currently at index $j$ costs $j$. In the *partial cost model*, the final positive access is not counted, so accessing an item currently at index $j$ costs $j-1$, denoted negative accesses.

---

[1] In the many papers that discuss Randomized-Move-To-Front, we have not been able to find a reference to the paper with the first definition of the algorithm. However, [7] cites personal communication with J. Westbrook from 1996 regarding properties of the algorithm.

After accessing an item, it can be moved to any position further towards the front of the list without any additional cost. Such a move can be seen as a number of transpositions of the accessed item with items preceding it in the list. The transpositions used to perform such a move are denoted *free*. Furthermore, at any time, an algorithm may exchange two adjacent items in the list at a cost of one. Such a transposition is denoted a *paid* transposition. The objective of a list accessing algorithm is to use free and paid transpositions in order to minimize the overall cost of serving the request sequence. Further discussion of the modelling issues can be found in [10].

Many different algorithms have been proposed for the list accessing problem. Some of the most well-known deterministic paging algorithms are the following.

MTF (Move-To-Front): After accessing the requested item, MTF moves the item to the front of the list.

FC (Frequency-Count): After accessing the requested item, FC moves the item forward in the list such that the resulting list is in sorted order with respect to the frequency with which the items have been accessed, i.e., for every item, FC maintains a counter which is incremented on an access to the item and the list is sorted in non-increasing order of the counters. FC only moves the accessed item forward the least number of positions necessary to maintain the sorted order.

TS (Time-Stamp): After accessing item $a_i$, it is inserted in front of the first item $a_j$ (from the front of the list) that precedes $a_i$ in the list and was accessed at most once since the last access to $a_i$. The algorithm does nothing if there is no such item $a_j$ or if $a_i$ is accessed for the first time.

TRANS (Transpose): After accessing the requested item, it is transposed with the item in front of it in the list. If the item is already at the front of the list, it stays there.

In addition to the above deterministic algorithms, we also consider the following well-known randomized algorithms.

BIT: For each item in the list, BIT [27] maintains a bit. Before processing a request sequence, BIT initializes the bits independently and uniformly at random. On a request for an item, BIT first complements the items bit. If the bit is then one, the item is moved to the front of the list. Otherwise, BIT does not move the item.

RMTF (Randomized-Move-To-Front): After accessing the requested item, RMTF moves the item to the front of the list with probability $\frac{1}{2}$.

## 3  Relative Worst Order Analysis

The relative worst order ratio was first introduced in [11] in an effort to combine the desirable properties of the max/max ratio [8] and the random-order ratio [23]. The measure was later refined in [12].

Instead of comparing online algorithms to an optimal offline algorithm (and then comparing their competitive ratios), two online algorithms are compared directly. However, instead of comparing their performance on the exact same request sequence, they are compared on their respective worst permutations of the same sequence.

Formally, if I is a request sequence of length $n$ and $\sigma$ is a permutation on $n$ elements, then $\sigma(I)$ denotes I permuted by $\sigma$. Let A be a list accessing algorithm and let A(I) denote the cost of running

A on I. Define $A_W(I)$ to be the performance of A on a worst possible permutation of I with respect to A, i.e., $A_W(I) = \max_\sigma \{A(\sigma(I))\}$.

For any pair of algorithms A and B, we define

$$c_u(A, B) = \inf\{c \mid \exists b \colon \forall I \colon A_W(I) \leq c\,B_W(I) + b\} \quad \text{and}$$
$$c_l(A, B) = \sup\{c \mid \exists b \colon \forall I \colon A_W(I) \geq c\,B_W(I) - b\}.$$

Intuitively, $c_l$ and $c_u$ can be thought of as tight lower and upper bounds, respectively, on the performance of A relative to B.

If $c_l(A, B) \geq 1$ or $c_u(A, B) \leq 1$, the algorithms are said to be *comparable* and the *relative worst order ratio*, $WR_{A,B}$, of algorithm A to algorithm B is defined. Otherwise, $WR_{A,B}$ is undefined.

$$\text{If } c_u(A, B) \leq 1, \text{ then } WR_{A,B} = c_l(A, B), \text{ and}$$
$$\text{if } c_l(A, B) \geq 1, \text{ then } WR_{A,B} = c_u(A, B).$$

When either $c_l(A, B) \geq 1$ or $c_u(A, B) \leq 1$ holds, the relative worst order ratio is a bound on how much better the one algorithm can be. If $WR_{A,B} < 1$, then A is better than B, and if $WR_{A,B} > 1$, then B is better than A. If A is better than B according to the relative worst order ratio, A and B are said to be comparable in A's favor. Finally, if the ratio is one, then the algorithms perform identically according to the relative worst order ratio.

In [11,12], it was shown that the relative worst order ratio is a transitive measure, i.e., the relative worst order ratio defines a partial ordering of the algorithms for a given problem.

## 4 List Factoring

The list factoring technique was first introduced by Bentley and McGeoch [9] and later extended and improved in a series of papers [21,29,1,3]. It reduces the analysis of list accessing algorithms to lists of size two. Previously the technique was developed and applied only in the context of competitive analysis, where it can be used to prove upper bounds on the competitive ratio [10]. In this section, we show that list factoring can also be applied in the context of relative worst order analysis to separate online algorithms and prove upper bounds.

In the following, let A denote any online list accessing algorithm that does not use paid transpositions. We are going to consider the partial cost model where accessing the $i$th item in the list costs $i - 1$. For any request sequence I, let $A^\star(I)$ denote the cost A incurs while processing I in the partial cost model.

Consider the list when A is about to process the $i$th request $I_i$ and define

$$A^\star(a_j, i) = \begin{cases} 1 & \text{if } a_j \text{ is in front of } I_i \text{ in the list} \\ 0 & \text{otherwise (including } a_j = I_i) \end{cases}$$

for all items $a_j$ in the list.

We also define

$$A^\star_{ab}(I) = \sum_{i : I_i \in \{a,b\}} (A^\star(a, i) + A^\star(b, i))$$

It is an easy observation, also made in [10], that we can then write the cost of A on a sequence I in the partial cost model as

$$A^\star(I) = \sum_{\{a,b\} \subseteq \mathcal{L}, a \neq b} A^\star_{ab}(I)$$

Let $I_{ab}$ be the projection of I over $a$ and $b$, i.e., the sequence obtained from I by deleting all requests to items different from $a$ or $b$.

An algorithm A is said to have the *pairwise property*, if for all pairs, $a$ and $b$, of two items in $\mathcal{L}$, we have
$$A_{ab}^\star(I) = A^\star(I_{ab})$$

In competitive analysis, this setup can be used to prove upper bounds on the competitive ratio (in the partial cost model) of algorithms that have the pairwise property. In addition, if the algorithms also are cost independent (the decisions they make are independent of the cost), then the ratio carries over to the full cost model [10].

For the relative worst order ratio, we show that this technique can also be used to separate algorithms and prove upper bounds.

Consider an algorithm A that has the pairwise property. It follows that
$$A_W^\star(I_{ab}) = \max_\sigma A^\star(\sigma(I_{ab})) = \max_\sigma A^\star((\sigma(I))_{ab}) = \max_\sigma A_{ab}^\star(\sigma(I))$$

The three equalities follow from the definition of a worst order, simple properties of permutations, and the pairwise property, respectively.

We now say that A has the *worst order projection property*, if and only if for all sequences I, there exist a worst ordering $\sigma_A(I)$ of I with respect to A, such that for all pairs $\{a,b\} \subseteq \mathcal{L}$ ($a \neq b$), $\sigma_A(I)_{ab}$ is a worst ordering of $I_{ab}$ with respect to A.

Using the above, we obtain a lemma similar to the Factoring Lemma for competitive analysis [10].

**Lemma 1.** *Let* A *and* B *be two online list accessing algorithms that do not use paid transpositions and that have the pairwise property and the worst order projection property, and let $\mathcal{L}$ be a list. If there exists constants $c$ and $b_1$ such that for every pair $\{a,b\} \subseteq \mathcal{L}$ ($a \neq b$), and for every request sequence I, $A_W^\star(I_{ab}) \leq c\,B_W^\star(I_{ab}) + b_1$, then there exists a constant $b_2$ such that for every request sequence I, $A_W^\star(I) \leq c\,B_W^\star(I) + b_2$.*

*In addition, if* A *and* B *are cost independent and $c \geq 1$, then $A_W(I) \leq c\,B_W(I) + b_2$.*

*Proof.* Consider any algorithm A satisfying the hypothesis. Then $A_W^\star(I)$ equals
$$\max_\sigma A^\star(\sigma(I)) = \max_\sigma \sum_{\{a,b\}\subseteq\mathcal{L},a\neq b} A_{ab}^\star(\sigma(I)) = \sum_{\{a,b\}\subseteq\mathcal{L},a\neq b} \max_\sigma A_{ab}^\star(\sigma(I)) = \sum_{\{a,b\}\subseteq\mathcal{L},a\neq b} A_W^\star(I_{ab})$$

Now consider two algorithms A and B satisfying the hypothesis. We get
$$A_W^\star(I) = \sum_{\{a,b\}\subseteq\mathcal{L},a\neq b} A_W^\star(I_{ab}) \leq \sum_{\{a,b\}\subseteq\mathcal{L},a\neq b} (c\,B_W^\star(I_{ab}) + b_1)$$
$$= c \sum_{\{a,b\}\subseteq\mathcal{L},a\neq b} B_W^\star(I_{ab}) + \sum_{\{a,b\}\subseteq\mathcal{L},a\neq b} b_1 = c\,B_W^\star(I) + \binom{\ell}{2} b_1$$

Hence, we have the result in the partial cost model. Now assume A and B are cost independent and $c \geq 1$. It is clear that for a cost independent algorithm A, the cost in the partial and the full cost model are related as $A_W(I) = A_W^\star(I) + |I|$. Hence, $A_W^\star(I) \leq c\,B_W^\star(I) + b$ implies that $A_W(I) \leq c\,B_W(I) + b$ and the result follows.

It follows from the above that we can use list factoring to separate online algorithms, and an upper bound on the relative worst order ratio on lists of size two carries over to lists of any size. However, as it is also the case for competitive analysis, the list factoring technique cannot be used to prove lower bounds.

For *randomized algorithms*, the worst ordering is defined in terms of the algorithm's excepted cost when run on the sequence. In this case, a randomized algorithm is said to have either of the two properties if for all settings of the random choices made by the algorithm (a deterministic execution of the algorithm), the property holds. With this definition, it is clear that the list factoring technique can also be applied to randomized algorithms.

In the following, we repeatedly use the fact that MTF, FC, and TS have the pairwise property and are cost independent [10].

## 5   Worst Orderings

Intuition suggests that one can obtain a worst ordering of any sequence for most online list accessing algorithms by considering the request sequence as a multiset of items and always request the item from the multiset which currently is farthest back in the list.

Formally, for any deterministic online list accessing algorithm A and any request sequence I, we inductively define the *FB ordering* (Farthest Back ordering) of I as follows. Let $S^0$ be the multiset of all items requested in I. Let $S^{i-1}$ be $S^0$ with the first $i-1$ items in the FB ordering removed. The $i$th item in the FB ordering of I with respect to A, $\text{FB}_A(I)_i$, is the item in $S^{i-1}$ which currently is farthest back in the list after A has processed the first $i-1$ requests of $\text{FB}_A(I)$. In addition, we say that A has the *FB property* if for any request sequence the FB ordering of that sequence is a worst ordering with respect to A.

When the algorithm in question is obvious, we drop it from the notation and write FB(I). Note that for any deterministic algorithm and request sequence, the FB ordering of this input sequence is uniquely determined.

Observe that TRANS does not have the FB property as the following example illustrates. Consider the request sequence $I = \langle a, b, c, c \rangle$ with the initial list $\mathcal{L} = (a, b, c)$. In this case, we have $\text{FB}(I) = \langle c, b, c, a \rangle$ with $\text{TRANS}(\text{FB}(I)) = 10$. However, on the ordering $I' = \langle c, c, b, a \rangle$, TRANS incurs a cost of 11. Hence, FB(I) is not a worst ordering for TRANS.

The other deterministic algorithms considered in this chapter do have the FB property.

**Lemma 2.** *MTF has the FB property.*

*Proof.* Consider any request sequence and let I be a worst ordering of this sequence with respect to MTF. We gradually reorder this sequence into the FB ordering maintaining at least the same cost, thereby proving the result. At each step we increase the length of the FB ordered prefix by at least one request. Hence, the process terminates after a finite number of steps.

Consider the first request $I_i$ in I, which differs from the FB ordering of I, i.e., $I_i = a \neq b = \text{FB}(I)_i$. Let $I_j$ be the first request to $b$ in I after $I_i$. If $b$ is also requested later than $I_j$ in I, let $I_k$ be this next request. Otherwise, let $I_k$ denote the last request in I.

Reorder I into I' by moving $I_j = b$ just in front of $I_i = a$, i.e.,

$$I = \langle \cdots, I_{i-1},\ \ a, I_{i+1}, \cdots, I_{j-1}, b, I_{j+1}, \cdots \rangle$$
$$I' = \langle \cdots, I_{i-1}, b, a, I_{i+1}, \cdots, I_{j-1},\ \ I_{j+1}, \cdots \rangle$$

First note that since MTF has the pairwise property, moving $b$ in the request sequence only affects $b$'s position in the list, the relative order of all other items remaining the same. Hence, if $b$ is accessed more than once after $I_j$ in I, then after the second access to $b$ (at $I_k$), the list is ordered the same for I and I$'$. Consequently, we only need to consider items requested in the subsequence $\langle I_i, \ldots, I_k \rangle$ to prove that the cost of I$'$ is at least the cost of I. Again, since MTF has the pairwise property, for each of these items, we only need to consider the number of negative comparisons between $b$ and the item accessed in both sequences.

Let $d$ be any item requested in $\langle I_i, \ldots, I_k \rangle$, $d \neq b$. In the following, the pair $(n, m)$ denotes that in I, $d$ is requested $n$ times in $\langle I_i, \ldots, I_{j-1} \rangle$ and $m$ times in $\langle I_{j+1}, \ldots, I_k \rangle$. We now have several cases depending on $n$ and $m$.

First, consider the pair $(0, m)$, $m \geq 0$. Since $d$ is not accessed in $\langle I_i, \ldots, I_{j-1} \rangle$, the relative order of $b$ and $d$ in the list is the same whenever they are requested in I or I$'$. Consequently, we have the same number of negative comparisons between $b$ and $d$ in the two sequences.

Next, consider the pair $(n, 0)$, $n > 0$. For both I and I$'$, we have a negative comparison with $d$ at the first access to $b$, since $b$ is at the end of the list. For I, there are no negative comparisons at the $n$ accesses to $d$. If $I_k$ is an access to $b$, then there is one negative comparison at this access. For I$'$, since $b$ is moved to the front of the list at its first access, there is one negative comparison at the first of the $n$ accesses to $d$. Also, if $I_k$ is an access to $b$, there is always one negative comparison at this access. Hence, in this case the number of negative comparisons always increase.

Now consider the pair $(n, m)$, $n > 0, m > 0$. Again, for both I and I$'$, we have one negative comparison at the first access to $b$. For I, since $d$ is before $b$ in the list just after $I_{i-1}$, there are no negative comparisons at the first $n$ accesses to $d$. There is one negative comparison at the first of the $m$ accesses to $d$, and one at the access $I_k$ if it is an access to $b$. For I$'$, there is one negative comparison at the first of the $n + m$ accesses to $d$, and a negative comparison at the access $I_k$ if it is an access to $b$. Hence, in this case the number of negative comparisons is at least the same in I$'$ as in I.

In all cases, the cost for MTF when processing I$'$ with respect to $b$ and $d$ is at least the same as the cost when processing I. □

**Lemma 3.** TS *has the* FB *property.*

*Proof.* Consider any request sequence and let I be a worst ordering of this sequence with respect to TS. We gradually reorder this sequence into the FB ordering maintaining at least the same cost, thereby proving the result. At each step we increase the length of the FB ordered prefix by at least one request. Hence, the process terminates after a finite number of steps.

Observe that for TS and any input sequence, the ordering of the items in the list at any point in time only depends on the initial ordering of the items and the last two accesses to each item. By this observation and the definition of TS, it follows by induction that the FB ordering of a request sequence repeatedly accesses the same item twice in a row (possibly except for the last access to any item), i.e., the item farthest back in the list. Note that the second access moves the item to the front of the list.

Divide FB(I) into phases corresponding to these pairs, i.e., a phase has length one or two.

Consider the first request $I_i$ in I which differs from the FB ordering, i.e., FB(I)$_i = a \neq I_i$. Hence, $a$ is the item farthest back of the items with remaining requests.

First, assume that FB(I)$_i$ is the only request in its phase, i.e., this is the last access to $a$ in FB(I). By the observation above, $a$ does not change its position in the list. It follows that the cost of this access is the same, independently of when it is made, i.e., we maintain the same cost by moving the one remaining request for $a$ in I forward to just before $I_i$ and as a consequence increase the prefix which is identical with the FB ordering.

Now, assume that $\mathrm{FB(I)}_i$ is the first of two requests to $a$ in its phase. Further, assume that at least two other requests to $a$ remains in FB(I) after this phase. We reorder I in the following way ($a_i$, $1 \leq i \leq 4$, are the first four of the remaining requests to $a$).

$$\mathrm{I} = \langle \cdots, \overbrace{I_i, \cdots}^{A}, a_1, \overbrace{\cdots}^{B}, a_2, \overbrace{\cdots}^{C}, a_3, \overbrace{\cdots}^{D}, a_4, \cdots \rangle$$

$$\mathrm{I'} = \langle \cdots, a_1, a_2, \overbrace{I_i, \cdots}^{A}, \overbrace{\cdots}^{B}, \overbrace{\cdots}^{C}, a_3, \overbrace{\cdots}^{D}, a_4, \cdots \rangle$$

We need to show that the cost for TS to serve I$'$ is at least as high as the cost for serving I. By the observation about TS, the behavior of TS after $a_4$ is the same for both sequences. Consider any item $d \neq a$ which is requested in either $A, B, C$, or $D$. Since TS has the pairwise property, we only need to show that the number of negative comparisons between $d$ and $a$ has not decreased. Again, it follows from the observation that we can assume $d$ is requested $0, 1$, or $2$ times in each subsequence.

Recall that $a$ does not change its position as the last item in the list when accessed at $a_1$ (in both sequences). Hence, for both sequences, there are negative comparisons between $a$ and $d$ at the accesses $a_1$ and $a_2$. In I, there are no negative comparisons in $A$ and $B$.

First assume that $d$ is requested two or more times in total in $A$, $B$, and $C$. In I$'$, we have two negative comparisons in total in $A$, $B$, and $C$ and one at each of $a_3$ and $a_4$ for a total of four negative comparison between $a$ and $d$. Now consider I. If $d$ occurs two or more times in $C$, we have two negative comparisons in $C$ and one at both $a_3$ and $a_4$ (no negative comparisons in $D$ since $a$ is not moved in front of $d$ at $a_3$). If $d$ occurs only once in $C$, we have at most one negative comparison in $C$, one at $a_3$, at most one in $D$, and at most one negative comparison at $a_4$. Finally, if $d$ does not occur in $C$, we have at most one negative comparison at $a_3$, two in $D$, and one at $a_4$. In all cases, we have at most four negative comparisons. Hence, the cost for TS to serve I$'$ is at least as high as the cost for serving I.

Next, assume that $d$ is requested once in total in $A$, $B$, and $C$. In I$'$, we have one negative comparison in total in $A$, $B$, and $C$, and if $d$ occurs in $D$, we have one negative comparison in $D$ and one at $a_4$. Now consider I. As noted above, if the access to $d$ occurs in either $A$ or $B$, then we have no negative comparisons in total in $A$, $B$, and $C$. If the access to $d$ occurs in $C$, then we have one negative comparison at that access. If $d$ occurs in $D$, we have one negative comparison in $D$ and one at $a_4$. In all cases, we have at least the same number of negative comparisons in I$'$ as in I.

As the last case, assume that $d$ does not occur in $A$, $B$, and $C$. It is clear that from the perspective of $a$ and $d$, the two sequences I and I$'$ are identical. Hence, they have the same number of negative comparisons between $a$ and $d$.

Finally, observe that the arguments above still hold in the case where there are one or no further requests for $a$ after $a_2$, i.e., $a_4$ or both $a_4$ and $a_3$ do not exist. Similarly, if $\mathrm{FB(I)}_i$ is not the first but the second of the two requests for $a$ in its subphase, the arguments still hold (this case corresponds to $A$ being the empty sequence).

**Lemma 4.** *FC has the FB property.*

*Proof.* Consider any request sequence and let I be a worst ordering of this sequence with respect to FC. We gradually reorder this sequence into the FB ordering maintaining at least the same cost, thereby proving the result. At each step we increase the length of the FB ordered prefix by at least one request. Hence, the process terminates after a finite number of steps.

Consider the first item in I which differs from the FB ordering of I, i.e., $I_i \neq \text{FB}(I)_i = b$. Let $I_j$ be the first request to $b$ in I after $I_i$. Reorder I into I′ by swapping $I_j$ with the preceding item, i.e., with the item at position $I_{j-1} = c$.

$$I = \langle \cdots, I_{j-2}, c, b, I_{j+1}, \cdots \rangle$$
$$I' = \langle \cdots, I_{j-2}, b, c, I_{j+1}, \cdots \rangle$$

We need to show that the cost incurred by FC to serve I′ is at least the cost it incurs when serving I. By induction, we can then swap $b$ into place as the $i$th request without reducing the overall cost.

Since FC has the pairwise property, we only need to consider the relative positions of $b$ and $c$ and the number of negative comparisons between the two.

We have three cases.

First, assume that the frequency of $c$ just after $I_{j-2}$ is lower than the frequency of $b$. The frequency of $c$ just before $I_i$ was also lower than the frequency of $b$ at that point. Hence, by the FC policy $c$ was further back than $b$ in the list just before $I_i$. Thus, the next request in the FB ordering would not have been to $b$, and we have reached a contradiction with the assumption that $b$ is the $i$th request in the FB ordering.

Next, assume that the frequency of $c$ just after $I_{j-2}$ is equal to the frequency of $b$. Since $b$ is further back in the list than $c$, there is one extra negative comparison in I′ in comparison with I up until $I_{j+1}$. However, the relative ordering of $b$ and $c$ in the list is now reversed going from I to I′, which may later cause one fewer negative comparison in I′ in comparison with I. Overall, the number of negative comparisons have not decreased.

Finally, assume that the frequency of $c$ just after $I_{j-2}$ is higher than the frequency of $b$. In this case, $c$ stays in front of $b$ in the list for both sequences. Hence, the cost for FC to serve both sequences is exactly the same.

When applying the list factoring technique in the next section, we use the following lemma.

**Lemma 5.** *If a deterministic algorithm has the* FB *property, then it also has the worst order projection property.*

*Proof.* This follows directly since a projection of an FB ordering is again an FB ordering.

Thus, MTF, FC, and TS all have the worst order projection property.

## 6 Algorithm Comparisons

We now have the tools necessary to compare the online list accessing algorithms.

### 6.1 Deterministic Algorithms

**Theorem 1.** *The algorithms* MTF *and* FC *perform identically according to the relative worst order ratio.*

*Proof.* We apply the list factoring technique introduced in Section 4 since both FC and MTF have the FB property.

Consider any request sequence I and any pair $\{a,b\} \subseteq \mathcal{L}$, $a \neq b$. Assume without loss of generality that the initial list has $a$ in front of $b$, i.e., $\mathcal{L}_{ab} = (a,b)$.

Now, the FB ordering of $I_{ab}$ for MTF is of the form $\langle (b,a)^m \rangle$ with a possible tail of repeated requests to either $a$ or $b$, whichever is requested the most in I. The FB ordering for FC is of the form $\langle (b,a,a,b)^{\lfloor \frac{m}{2} \rfloor} \rangle$ with a possible tail of repeated requests to either $a$ or $b$, whichever is requested the most in I. Observe that if $m$ is not divisible by two, there is an extra request to either $a$ or $b$. However, such a request only contributes a constant extra cost which we can ignore. It now follows that the cost for FC on its worst permutation (the FB ordering) is the same as the cost for MTF on its worst permutation (the FB ordering), except for a possible additive constant.

**Theorem 2.** *The algorithms* MTF *and* TS *perform identically according to the relative worst order ratio.*

*Proof.* We apply the list factoring technique introduced in Section 4 since both TS and MTF have the FB property.

Consider any request sequence I and any pair $\{a,b\} \subseteq \mathcal{L}$, $a \neq b$. Assume without loss of generality that the initial list projected onto $a$ and $b$ has $a$ at the front, i.e., $\mathcal{L}_{ab} = (a,b)$.

The FB ordering of $I_{ab}$ for TS is of the form $\langle (b,b,a,a)^{\lfloor \frac{m}{2} \rfloor} \rangle$. The remaining arguments are exactly the same as in the proof of Theorem 1.

Combining the previous two lemmas and using the fact that the relative worst order ratio is a transitive measure, we arrive at the following corollary.

**Corollary 1.** *The algorithms* MTF, TS, *and* FC *perform identically according to the relative worst order ratio.*

We now show that TRANS cannot be better than any of MTF, TS, and FC according to the relative worst order ratio.

**Lemma 6.** *There exists a constant b such that for any request sequence* I,

$$\mathrm{MTF}_W(I) \leq \mathrm{TRANS}_W(I) + b$$

*Proof.* Consider any request sequence I. Reorder I into an FB ordering with respect to MTF, $I_{\mathrm{MTF}}$, and recursively divide it into phases as follows. Each phase starts where the previous phase ends. Let $n_d$ be the number of distinct item in the remaining part of the sequence (the part which has not yet been divided into phases). The next phase contains the next $n_d$ requests in the sequence. Since $I_{\mathrm{MTF}}$ is an FB ordering with respect to MTF and by the MTF policy, all accesses in a phase are to distinct items. In addition, the relative order of the items accessed in a phase is the same immediately before and after the phase.

Now group the phases into super phases, where each super phase is a maximal sequence of consecutive phases under the restriction that the phases request the same number of distinct items. Hence, each of the accesses in each phase of a super phase are to the same set of items. Let $r_i$ denote the number of phases in the $i$th super phase, and let $n_i$ denote the number of distinct items accessed in each phase of the $i$th super phase. For convenience, we define $n_0 = \ell$, recalling that $\ell$ is the length of the list $\mathcal{L}$.

Observe that the number of distinct items in the super phases are decreasing and the number of distinct items in the first phase is at most $\ell$. Hence, there are at most $\ell$ super phases.

For MTF, the cost of the $i$th super phase can now be calculated as follows. The first phase of the $i$th super phase costs at most $n_{i-1}n_i$, since the $n_i$ accesses in the worst case is for items at index $n_{i-1}$ in the list (by the MTF policy). The remaining $r_i - 1$ phases in the super phase cost $n_i^2$ each. Hence, the total cost for MTF of the $i$th super phase is $n_i(n_{i-1} - n_i) + r_i n_i^2$. The only term we are interested in is $r_i n_i^2$, since the remaining terms, over all super phases, can be bounded by a constant depending only on $\ell$. Thus, this constant is independent of the length of the request sequence.

Hence, we need to show that we can find an ordering making TRANS incur a cost of at least $r_i n_i^2$, up to an additive constant.

We ignore super phases with $r_i \leq 2\ell$. The cost for MTF on such phases is at most $r_i n_i \ell \leq 2\ell^3$. Since there are at most $\ell$ phases, the total cost incurred by MTF on these super phases is only a constant dependent on $\ell$.

Now, assume that the $i$th super phase has $r_i > 2\ell$. We reorder the requests for TRANS. First, we access each of the $n_i$ items $\ell$ times, which moves the $n_i$ items to the first $n_i$ positions in the list. Assume that the first $n_i$ items in the list after this are $(a_1, a_2, \ldots, a_{n_i})$ where $a_1$ is at the front of the list.

First, if $n_i$ is one, simply repeatably access the item, giving a total cost of at least $r_i = r_i n_i^2$.

Next, if $n_i$ is even, let $r_i' = r_i - \ell - n_i > 0$. Access $a_{n_i}$ and $a_{n_{i-1}}$ alternately $r_i'$ times with a cost of $n_i$ for each access. Subsequently, access $a_{n_i}$ $n_i$ times, and then $a_{n_{i-1}}$ $n_i$ times, thereby moving them to the front of the list. Repeat this process for the remaining $n_i - 2$ items in groups of two. The total cost is at least $(\ell + r_i' n_i + n_i) n_i = \ell n_i + r_i n_i^2 - \ell n_i^2 - n_i^3 + n_i^2$.

Finally, if $n_i$ is odd and at least three, we do as in the previous case, except when we are down to the last three items, $a_1$, $a_2$, and $a_3$ at positions $n_i$, $n_{i-1}$, and $n_{i-2}$. Let $r_i'' = \lfloor \frac{r_i - \ell - 2}{2} \rfloor$ and request $\langle (a_1, a_2)^{r_i''}, a_1, a_1, (a_2, a_3)^{r_i''}, a_2, a_2, (a_3, a_1)^{r_i''}, a_3, a_3 \rangle$. All requests have a cost of $n_i$ except for the last of the double accesses to $a_1$, $a_2$, and $a_3$ each with a cost of $n_i - 1$. In total, the three items are each accessed $\ell + 2r_i'' + 2 \in \{r_i - 1, r_i\}$ times. If this is $r_i - 1$, request the items $\langle a_1, a_2, a_3 \rangle$, each at a cost of $n_i$. In total, the cost per item is $r_i n_i - 1$.

In both of the above two cases, we can ignore all terms, except for terms involving $r_i$. Hence, the total cost for TRANS in this super phase is at least $r_i n_i^2$, except for a constant dependent only on $\ell$, i.e., the difference in cost for MTF and TRANS is bounded by a constant only dependent on $\ell$.

On the other hand, TRANS can be much worse than MTF, FC, and TS under the relative worst order ratio.

**Theorem 3.** $\text{WR}_{\text{TRANS,MTF}} \geq \frac{\ell}{2}$.

*Proof.* Lemma 6 shows that TRANS cannot be better than MTF according to the relative worst order ratio. Assume that the initial list is $\mathcal{L} = (a_1, a_2, \ldots, a_\ell)$ and consider the request sequence $I = \langle (a_\ell, a_{\ell-1})^m \rangle$.

It is clear that MTF incurs a cost of $2\ell + 4(m-1)$ on its worst permutation of I. On the other hand, TRANS leaves the two items at the end of the list and incurs a cost of $2m\ell$. For $m$ approaching infinity, the ratio approaches $\frac{\ell}{2}$.

### 6.2 Randomized Algorithms

In this section, to make the proofs more readable, we use the partial cost model. Here, as it is the case in the rest of this paper, all results hold for the full cost model as well.

**Lemma 7.** *For integers $n \geq 1$ and $m \geq 2$ and a request sequence $I = \langle (b, a^m)^n \rangle$ with initial list $\mathcal{L} = (a, b)$, the expected cost of BIT for a single repetition of $\langle b, a^m \rangle$ is $\frac{7}{4}$, and $I$ is its own worst permutation with respect to BIT.*

*Proof.* For each access to $b$, at most the next two accesses to $a$ contribute to the expected cost of BIT. It follows by induction that after each repetition of $\langle b, a^m \rangle$, $a$ is at the front of the list for BIT. Hence, the expected cost of the prefix $\langle b, a \rangle$ of the next repetition is $\frac{3}{2}$, and after that $a$ is at the front of BIT's list with probability $\frac{3}{4}$. Thus, the expected cost of the following access to $a$ is $\frac{1}{4}$, after which $a$ is at the front of BIT's list with probability 1, and the remaining accesses to $a$ in the current repetition do not cost anything. Hence, for any $m$, the total expected cost of a single repetition is $\frac{7}{4}$ for BIT. It is clear that I is its own worst permutation for BIT.

**Lemma 8.** *There exists a request sequence $I$ such that the expected cost for RMTF on its worst permutation of $I$ is strictly less than the expected cost for BIT on its worst permutation.*

*Proof.* Consider the request sequence $I = \langle (b, a, a)^n \rangle$ for some integer $n$ with the initial list $\mathcal{L} = (a, b)$.

For BIT, by Lemma 7, the cost of each repetition of $\langle b, a, a \rangle$ is $\frac{7}{4}$.

For RMTF, first consider any subsequence $\langle b, a^m \rangle$ of a worst ordering of I for some positive integer $m$. Assume that before this subsequence, in RMTF's execution, $a$ is at the front of the list with probability $p$.

After the access to $b$, $a$ is not at the front with probability $1 - \frac{p}{2}$. In this case, the up to $m$ requests to $a$ while it is not at the front can be described by a truncated geometric distribution [10, Lemma 4.1] with an expected number of $2\left(1 - \frac{1}{2^m}\right)$. Hence, the cost of the entire subsequence is

$$c_m(p) = p + \left(1 - \frac{p}{2}\right) 2 \left(1 - \frac{1}{2^m}\right) = 2 - \frac{2-p}{2^m}$$

The probability of $a$ being at the front of the list after the repetition is then

$$1 - \frac{1 - \frac{p}{2}}{2^m} = 1 - \frac{2-p}{2^{m+1}}$$

Now, consider the input sequence $\langle (b, a^m)^n \rangle$ for $n$ approaching infinity. The probability of $a$ being at front of the list after each repetition of $ba^m$ approaches $p_m$, where

$$p_m = 1 - \frac{2 - p_m}{2^{m+1}} \Rightarrow p_m = 1 - \frac{1}{2^{m+1} - 1}$$

Hence, the cost of a repetition approaches

$$c_m(p_m) = 2 - \frac{2 - p_m}{2^m} = 2 - \frac{1}{2^m} - \frac{1}{2^m(2^{m+1} - 1)} = \frac{2^{m+2} - 4}{2^{m+1} - 1}$$

The results are summarized in Table 1, including values for small $m$.

| $m$ | $c_m(p)$ | $p$ after phase | $p_m$ | $c_m(p_m)$ |
|---|---|---|---|---|
| 0 | $p$ | $\frac{p}{2}$ | 0 | 0 |
| 1 | $1 + \frac{p}{2}$ | $\frac{1}{2} + \frac{p}{4}$ | $\frac{2}{3}$ | $\frac{4}{3}$ |
| 2 | $\frac{3}{2} + \frac{p}{4}$ | $\frac{3}{4} + \frac{p}{8}$ | $\frac{14}{15}$ | $\frac{12}{7}$ |
| 3 | $\frac{7}{4} + \frac{p}{8}$ | $\frac{7}{8} + \frac{p}{16}$ | $\frac{30}{31}$ | $\frac{28}{15}$ |
| $\geq 4$ | $\leq 2$ | $\leq 1$ | $\leq 1$ | $\leq 2$ |

**Table 1.** The cost and the value of $p$ after the phase for various $m$.

Returning to I, assume for the moment that the ordering of I is indeed a worst ordering for RMTF. By the above, it is clear that for $p < 1$, the expected cost for RMTF to serve a repetition is strictly less than the expected cost for BIT, and only on the very first repetition is $p = 1$; all following repetitions have $p < 1$. Also, the cost of a repetition approaches $c_2(p_2) = \frac{12}{7}$ for $n$ approaching infinity. This is strictly less than the cost of BIT, $\frac{7}{4}$.

Now, we only need to show that the ordering of I is a worst ordering for RMTF. Consider any other ordering I' of I. Since $a$ is initially at the front of the list, we assume without loss of generality that the first request in I' is for $b$. Divide I' into phases of the form $\langle b, a^m \rangle$. Consider any such phase and assume $a$ is at the front of the list with probability $p$ just before the phase.

The idea is to match phases for values of $m$ different from 2 in I' against a corresponding number of phases with $m = 2$ in I.

Before doing this, we first compare the cost for RMTF serving I with $a$ initially at the front with probability 1 against the cost for RMTF serving I where it is initially in the stable state, $p_2$. This later enables us to compare the cost of the phases in I' with the cost of I, assuming that the stable state $p_2$ has been reached.

Consider a sequence of phases with $m = 2$. As a worst case assumption, we assume that the length of the sequence of phases is infinite. Starting with a probability of 1, the probability of $a$ being at the front of the list just before RMTF starts to serve phase $j$, $j \geq 0$, is,

$$\frac{3}{4} + \frac{1}{8} \cdot \frac{3}{4} + \frac{1}{8^2} \cdot \frac{3}{4} + \cdots + \frac{1}{8^{j-1}} \cdot \frac{3}{4} + \frac{1}{8^j} = \frac{3}{4} \left( \sum_{i=1}^{j-1} \frac{1}{8^i} \right) + \frac{1}{8^j} = \frac{3}{4} \cdot \frac{1 - \frac{1}{8^j}}{1 - \frac{1}{8}} + \frac{1}{8^j} = \frac{6}{7} + \frac{1}{7} \cdot \frac{1}{8^j}$$

Hence, the contribution of phase $j$ to the extra cost is

$$c_2 \left( \frac{6}{7} + \frac{1}{7} \cdot \frac{1}{8^j} \right) - c_2(p_2) = \frac{3}{2} + \frac{6}{28} + \frac{1}{28} \cdot \frac{1}{8^j} - \frac{12}{7} = \frac{1}{28} \cdot \frac{1}{8^j}$$

Summing all the contributions, we get the extra cost of the sequence of phases beginning with $p = 1$ in comparison with starting in the stable state as

$$\sum_{j=0}^{\infty} \frac{1}{28} \cdot \frac{1}{8^j} = \frac{1}{28} \cdot \frac{1}{1 - \frac{1}{8}} = \frac{2}{49}$$

We now start to compare the sequence I' to I. We assume that all phases with $m \neq 2$ in I' begin with $a$ being at the front of the list with probability 1. We match phases in I' for values of $m \neq 2$ against a number of phases in I (which have $m = 2$), such that the number of $a$'s and $b$'s correspond. Observe that a phase in I$'$ might be followed by a sequence of phases with $m = 2$. If we start processing such a following sequence of phases with $m = 2$ and with $p > \frac{6}{7}$, then the cost of the sequence will be higher than if we started in the stable state. However, we have upper bounded the extra cost by $\frac{2}{49}$. Now, only phases with $m \geq 3$ can end with $p > \frac{6}{7}$ (see Table 1). Hence, in the following matching of phases, we add a contribution of $\frac{2}{49}$ to the cost whenever we match a phase with $m \geq 3$.

Consider the phases in I$'$. We repeatedly apply the following matching of phases until there are no phases or only phases with $m = 2$ are left, which we have accounted for by adding a contribution of $\frac{2}{49}$ to the possibly preceding phase.

- If there is an unmatched phase with $m = 3$ and an unmatched phase with $m = 1$ left in I$'$, then they correspond to exactly two phases with $m = 2$. An upper bound on the cost for RMTF when serving the requests in I$'$ is then $\left(\frac{7}{4} + \frac{1}{8}\right) + \left(1 + \frac{1}{2}\right) + \frac{2}{49}$, which is strictly less than the cost of the corresponding two phases in I when RMTF is in the stable state, $2 \cdot \frac{12}{7}$.
- If there are two unmatched phases with $m = 3$ and an unmatched phase with $m = 0$ left in I$'$, then they correspond to exactly three phases with $m = 2$. An upper bound on the cost for RMTF when serving the requests in I$'$ is then $2\left(\frac{7}{4} + \frac{1}{8}\right) + 1 + 2 \cdot \frac{2}{49}$, which is strictly less than the cost of the corresponding three phases in I when RMTF is in the stable state, $3 \cdot \frac{12}{7}$.
- If there are still phases with $m = 3$ left in I$'$ after applying the above two cases (repeatedly), then it follows that there must exist at least one phase with an odd $m > 3$ and a number of phases with $m = 0$ such that resulting number of $a$'s in the phases is two times the number of phases. Let $x$ denote the value of $m$ in the unmatched phase in I$'$ with the smallest value of $m$ under the restriction that $m$ is odd and $m \geq 5$. Then there are at least $\frac{3+x}{2} - 2 = \frac{x-1}{2}$ phases with $m = 0$. These phases in I$'$ correspond to $\frac{3+x}{2}$ phases with $m = 2$ in I. An upper bound on the cost for RMTF when serving the requests in I$'$ is then $\left(\frac{7}{4} + \frac{1}{8}\right) + 2 + \frac{x-1}{2} \cdot 1 + 2 \cdot \frac{2}{49} = \frac{x}{2} + \frac{1355}{392}$, which is strictly less than the cost of the corresponding phases in I when RMTF is in the stable state, $\frac{3+x}{2} \cdot \frac{12}{7} = \frac{6}{7}x + \frac{18}{7}$, for $x \geq 5$.
- If there is an unmatched phase with $m = 4$ and two unmatched phases with $m = 1$ left in I$'$, then they correspond to three phases with $m = 2$. An upper bound on the cost for RMTF when serving the requests in I$'$ is then $2 + 2\left(1 + \frac{1}{2}\right) + \frac{2}{49}$, which is strictly less than the cost of the corresponding three phases in I when RMTF is in the stable state, $3 \cdot \frac{12}{7}$.
- If there is an unmatched phase with $m = 4$ and an unmatched phase with $m = 0$ left in I$'$, then they correspond to two phases with $m = 2$. An upper bound on the cost for RMTF when serving the requests in I$'$ is then $2 + 1 + \frac{2}{49}$, which is strictly less than the cost of the corresponding two phases in I when RMTF is in the stable state, $2 \cdot \frac{12}{7}$.
- Finally, if there is an unmatched phase with $m > 4$, let $x$ denote the value of $m$ in the unmatched phase, and set $y = \lceil \frac{x}{2} \rceil \geq 3$. In I$'$, there must be $y - 1$ phases with $m \leq 1$. As a worst case assumption, we assume that all such phases have $m = 1$. Now, an upper bound on the cost for RMTF when serving the requests in I$'$ is then $2 + (y-1)\left(1 + \frac{1}{2}\right) + \frac{2}{49}$, which is strictly less than the cost of the corresponding $y$ phases in I when RMTF is in the stable state, $y\frac{12}{7}$, for $y \geq 3$.

Observe that the above covers all cases. Hence, the ordering in I is indeed a worst ordering for RMTF.

**Lemma 9.** *There exists a request sequence* I *such that the expected cost for* BIT *on its worst permutation of* I *is strictly less than the expected cost for* RMTF *on its worst permutation.*

*Proof.* Consider the request sequence $I = \langle (b, a, a, a)^n \rangle$, $n \geq 1$, with the initial list $\mathcal{L} = (a, b)$.

For BIT, by Lemma 7, the cost of each repetition of $\langle b, a, a, a \rangle$ is $\frac{7}{4}$, and I is its own worst permutation.

For RMTF, by Table 1, the cost of each repetition approaches $c_3(p_3) = \frac{28}{15}$ from above. Since this is strictly more than $\frac{7}{4}$, $\text{RMTF}(I) > \text{BIT}_W(I)$.

An interesting observation is that the sequences used in the previous lemmas are both repetitions of the pattern $\langle b, a^m \rangle$ for different values of $m$. The previous two lemmas imply the following:

**Corollary 2.** BIT *and* RMTF *are not comparable using the relative worst order ratio.*

## 7 Open Problems

In order to apply the list factoring technique together with relative worst order analysis, both of the "pairwise properties" and the "worst order projection properties" must hold. We have not been able to show a dependence between these two properties, i.e., does one follow from the other? On the other hand, we have not been able to exhibit an example for which one holds and the other does not.

Another interesting question is whether the list factoring technique can be used with performance measures other than competitive analysis and, as demonstrated here, relative worst order analysis.

## Acknowledgments

We would like to thank Joan Boyar for initial discussions on the relationship between MTF and TRANS.

16      M. R. Ehmsen, J. S. Kohrt, K. S. Larsen